\documentclass[a4paper,notitlepage,superscriptaddress,aps,pre,10pt,longbibliography]{revtex4-1}

\usepackage[T1]{fontenc}
\usepackage{float}
\usepackage{amsmath, amsthm, amssymb, amsfonts}
\usepackage{graphicx}
\usepackage[caption=false]{subfig}
\usepackage[english]{babel}

\begin{document}

\title{Sudden Trust Collapse in Networked Societies}

\author{Jo{\~a}o da Gama Batista}
\email{joao.batista@ecp.fr}
\affiliation{Chaire de Finance Quantitative, Laboratoire de Math{\'e}matiques Appliqu{\'e}es aux Syst{\`e}mes, {\'E}cole Centrale Paris, Ch{\^a}tenay-Malabry, 92290, France}

\author{Jean-Philippe Bouchaud}
\email{jean-philippe.bouchaud@cfm.fr}
\affiliation{Capital Fund Management, 23 rue de l'Universit{\'e}, 75007 Paris}

\author{Damien Challet}
\email{damien.challet@ecp.fr}
\affiliation{Chaire de Finance Quantitative, Laboratoire de Math{\'e}matiques Appliqu{\'e}es aux Syst{\`e}mes, {\'E}cole Centrale Paris, Ch{\^a}tenay-Malabry, 92290, France}
\affiliation{Encelade Capital SA, EPFL Innovation Park, B{\^a}timent C, 1015 Lausanne, Switzerland}

\begin{abstract}
	Trust is a collective, self-fulfilling phenomenon that suggests analogies with phase transitions. We introduce a stylized model for the build-up and collapse of trust in networks, which generically displays a first order transition. The basic assumption of our model is that whereas trustworthiness begets trustworthiness, panic also begets panic, in the sense that a small decrease in trustworthiness may be amplified and ultimately lead to a sudden and catastrophic drop of collective trust. We show, using both numerical simulations and mean-field analytic arguments, that there are extended regions of the parameter space where two equilibrium states coexist: a well-connected network  where global confidence is high, and a poorly connected network where global confidence is low. In these coexistence regions, spontaneous jumps from the well-connected state to the poorly connected state can occur, corresponding to a sudden collapse of trust that is not caused by any major external catastrophe. In large systems, spontaneous crises are replaced by {\it history dependence}: whether the system is found in one state or in the other essentially depends on initial conditions. Finally, we document a new phase, in which agents are well connected yet distrustful.
\end{abstract}
\maketitle

\section{Introduction\label{sec:intro}}

\begin{figure}[!htb]
	\centering
	\includegraphics[scale=0.45]{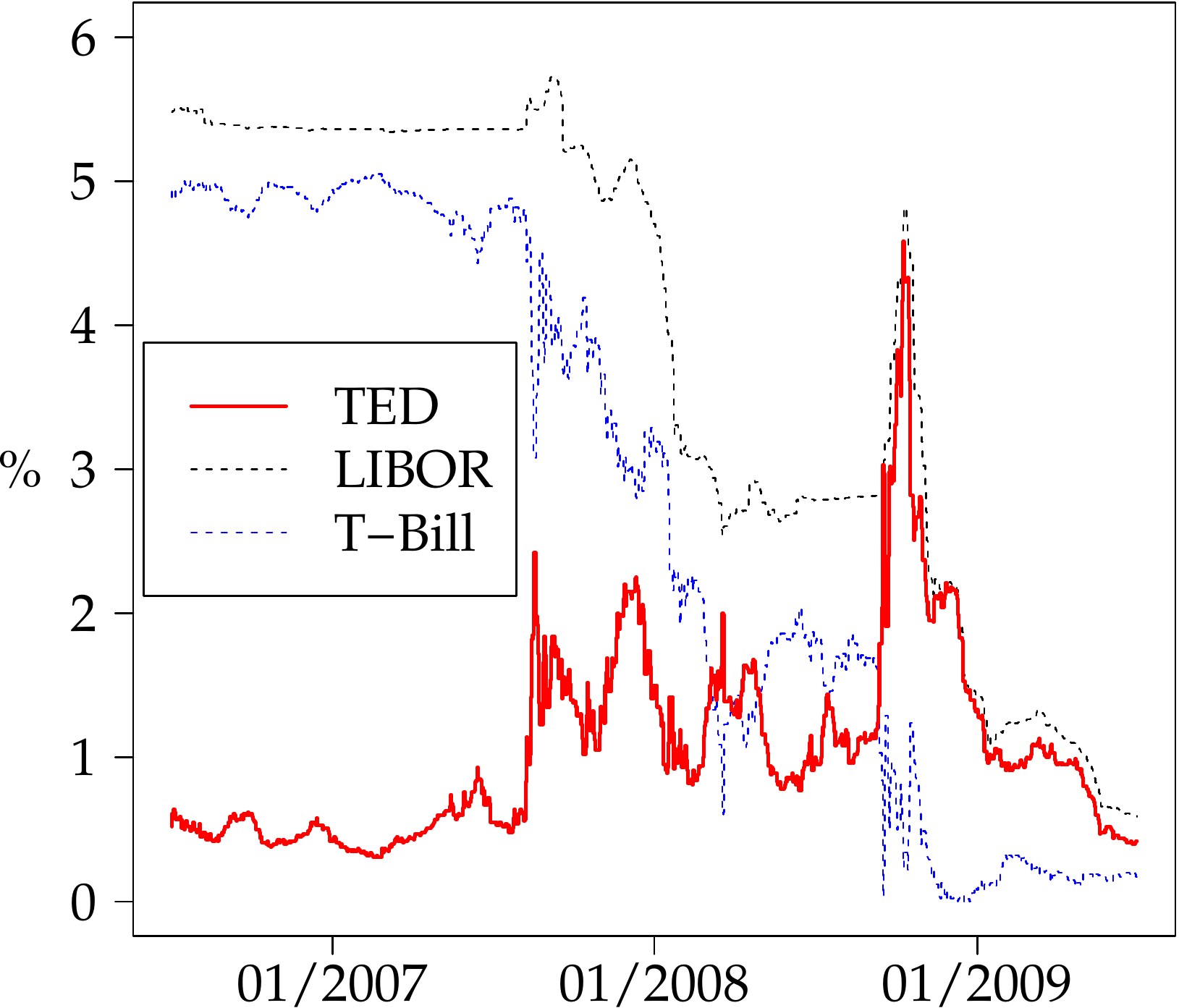}
	\caption[Content]{TED spread, three-month LIBOR and three-month T-bill interest rate (Jul. 2006 -- Jul.2009). The TED spread is the difference between the three-month LIBOR and the three-month T-bill interest rate. Taking into account that T-bills are considered risk-free, an increase in the TED spread is an indicator of higher perceived credit risk in the overall economy. In 10th October 2008, in the wake of the bankruptcy of Lehman Brothers, the TED spread reached $4.57\%$, several times above the long term average of $0.30\%$!}
	\label{fig:tedspread}
\end{figure}

In the wake of the 2008 crisis, President Barack Obama declared: {\it Our workers are no less productive than when this crisis began. Our minds are no less inventive, our goods and services no less needed than they were last week, or last month, or last year} \cite{Obama2009}. So what had happened that made the world so different from a few months before? No war or physical catastrophe had occurred that would have destroyed tangible assets, infrastructures or knowledge. As implied by President Obama's comment, the damage seems to have been, at least partially, self-inflicted by a sudden collapse of trust that led to a ``freeze'' of the interbank lending  network (evidenced by soaring interbank rates, see Fig. \ref{fig:tedspread}) and, nearly immediately afterwards, to a collapse of confidence of all economic actors -- investors, firms, households interrupted projects and reduced consumption, driving the economy to a grinding halt\footnote{Those who were in New York at the end of Sept. 2008 will remember the sight of completely empty retail stores and the stories of people emptying their bank accounts and going home with cash in plastic bags.}. The bewildering aspect of such a crisis (as well as 
many previous ones) is the speed at which financial markets, or the economy as a whole, can shift from a relatively efficient state to a completely dysfunctional one.  Whereas most ``real'' economic factors (technology, workforce, R\&D) usually change relatively slowly, trust or subjective expectations seem to have no inertia, no anchor to their past values, and can swing from high to low in a matter of days, hours or even minutes.

Trust is critical in determining the prosperity of human societies and to secure a well-functioning economy and orderly financial markets. Moreover, trust is a collective asset that allows efficient coordination and cooperation, and tremendously accelerates business. It allows for the emergence of genuinely collective figments, such as money and other social conventions. Fiat money is a perfect example: a piece of paper can only be valuable if everybody believes that it will not be worthless tomorrow, and if everybody does, bank notes indeed become valuable.

The fact that trust is (as we view it) a collective, self-fulfilling phenomenon suggests analogies with phase transition phenomena, where collective properties emerge that cannot exist at the individual level, like magnetism, superfluidity, etc. Magnets, for example, arise because the spin of each atom acquires a favoured orientation, imposed by the favoured orientation of neighbouring atoms. This occurs when the interaction between spins becomes strong enough. Clearly, trust emergence is similar, and follows from  positive feedback loops such as {\it I trust you because he trusts you because I trust you}. The most important aspect of the  analogy with phase transition is the possible coexistence of very different equilibrium states, which leads to dis-equilibrium phenomena like {\it history dependence} or ``hysteresis'', when the system is trapped in one equilibrium while another is more favourable, and discontinuities, when the system jumps from one state to the other. This is an interesting scenario as it opens a path to explain the sudden swings of trust that seem to underpin many economic, financial or political crises.

Several models for trust collapse have been studied along these lines in the past few years, see e.g. \cite{Anand2010,Anand2013,Cont2012,Amini2013,Caballero2009,Heise2012,Battiston2012,Harmon2010,Sieczka2011,Peixoto2012,Bouchaud2013,Gai2010,Gai2011,Caccioli2012,Contreras2014,Bianconi2004} and references therein. The common crucial feature is the coexistence of two (or more) equilibrium states in a region of the parameter space, and therefore the possibility of a sudden jump between a favourable, high-confidence state to an unfavourable, low-confidence state. In these models, the jump is not induced by a major catastrophe (that would {\it replace} the favourable equilibrium by an unfavourable one) but rather by some anecdotal random fluctuation, which can induce a transition toward an {\it already pre-existing} low-confidence equilibrium.

Here, we introduce and study a highly stylized model for the build-up and collapse of collective trust in a dynamically evolving network, which generically
displays a first order transition with possible coexistence of different equilibria. The nodes of the network can represent individuals, firms, banks, etc. Each node is assigned a real number that measures its (perceived) trustworthiness.

The presence of an undirected link between two nodes indicates an established relationship of some kind (business, loan, collaboration, etc.) resulting from some common rational benefit, but 
only possible if the perceived trustworthiness of the partner is high enough. Links are thus created or destroyed depending on the trustworthiness of the nodes and their dynamics; conversely, the trustworthiness of a node depends on that of its neighbours. The network and the trustworthiness therefore co-evolve and, depending on the precise specification of the model (see below), this leads to a rich dynamics with crises where the network disintegrates and the collective trust collapses. We solve our model within a mean-field approximation and find, as anticipated, that there is a region of parameters where different equilibria indeed coexist.

Our model and results are in several ways similar to those obtained by M. Marsili and associates in two very inspiring papers \cite{Marsili2004,Ehrhardt2006}. They also study the coupled dynamics of links and nodes and find generic phase coexistence and hysteresis. One new aspect of our work is to consider that the {\it speed of change} of trustworthiness is itself a piece of information which agents strongly react to, in particular when it is negative -- in a ``panic feeds panic'' spirit. Our mean-field analysis describes the phenomena induced by this effect and predicts phases which had not been considered before, such as a connected yet distrustful phase. In a sense, our model is a stylized version of \cite{Anand2010,Anand2013} that removes all the specifics of the interbank lending network, and a generalized version of \cite{Ehrhardt2006}, where some ingredients specific to the dynamics of trustworthiness are introduced, leading to new effects. The possible coexistence of different states has also been noted in the context of epidemic propagation on networks which may be rewired so as to avoid infected nodes. In this case, infected network situations may indeed coexist with healthy networks \cite{Pastor-Satorras2014}. This is similar to our model, where agents/firms/banks tend to cut their links with degraded nodes.

\section{The model}

\subsection{Trustworthiness of the nodes}

The nodes in the network are agents which can represent individuals, companies, banks or other institutions. We make the strong assumption that the {\it perceived} trustworthiness of a node $i$, which determines its propensity to link with other nodes, can be summarized by the value of a real number $-\infty < h_i < +\infty$. That real number may depend on a variety of factors, which can be deemed either objective or subjective depending on their underlying nature. The balance sheet of a bank or the health of a business are examples of objective or ``intrinsic'' factors. Subjective factors come into play, for instance, when one needs to assess how trustworthy the counterparties or business partners of $i$ are. Clearly, if the debtors of $i$ are close to bankruptcy, they endanger the balance sheet of $i$ itself -- this mechanism is at the core of many recent models of bankruptcy cascades such as \cite{Corsi2013,Sieczka2011,Contreras2014,Gai2010,Caccioli2012,Lorenz2009,Battiston2012}. But one can imagine different, less mechanical channels of propagation. A good example for our purpose is reputation risk. In fact, if node $j$ is caught up in a scandal while making business with $i$, other partners of $i$ might become wary that $i$ is also involved and decide to end their business with $i$, unless $i$ reacts immediately and severs its own link with $j$.

Another important factor is the speed of variation of the trustworthiness itself. Imagine a highly respected bank or institution $i$ that rapidly loses many of its partners. This will be interpreted as worrying news by the remaining partners who, as a precautionary measure, will be tempted to cut their relation as well, even if the trustworthiness of $i$ is still high. This ``bank run'' or ``panic'' type of feedback loop can be amplified by the existence of a CDS (Credit Default Swap) market, which is supposed to price the default probability of firms and banks (and countries) and thus a proxy for $h_i$. The very fact that the price of the CDS increases (and thus the perceived default probability) can trigger a crash-type dynamics.  These avalanches of sell-offs when the perceived risk increases are often observed in financial markets as a consequence of a highly conservative management of ``Black Swan'' events -- that, ironically, may result from these risk management policies!  

Mathematically, we therefore write the trustworthiness $h_{i}$ of each node $i$ as the sum of three terms:
\begin{equation}\label{h:def}
	h_{i}=h_{i,0}+ f h^* k_{i} \tanh \left(\frac{\overline{h_{i}}}{h^*}\right)+d\cdot \min \left(0,\ \delta h_i\right),\
\end{equation}
where $f,h^*,d$ are positive constants, $k_i$ is the degree of node $i$, $\overline{h_{i}} = (\sum_{j \in V_i} h_j)/k_i$ is the average trustworthiness of the nodes $j \in V_i$ that are connected to $i$ (with $\overline{h_{i}} \equiv 0$ if $k_i=0$), and $\delta h_i$ is the variation of $h_i$ over the last time step.

The first term $h_{i,0}$ is the intrinsic trustworthiness of node $i$, assumed here to be time-independent, IID random variables with mean $m$ and variance $\sigma^2$. More specifically we will choose $h_{i,0}$ to be uniformly distributed in the interval $[0,2]$, corresponding to a positive mean $m =1$ and $\sigma^2=\frac{1}{3}$.

The second term describes how much of the trustworthiness of the peers of $i$ is bequeathed to $i$. When $\overline{h_{i}}$ is much smaller than a characteristic value $h^*$, expanding $\tanh(x)$ for small arguments gives the following contribution:
\begin{equation}
	f h^* k_{i} \tanh \left(\frac{\overline{h_{i}}}{h^*}\right) \approx f \sum_{j \in V_i} h_j,
\end{equation}
which means that a fraction $f$ of the total trustworthiness of the business partners of $i$ is transferred to $i$ itself. The $\tanh$ function imposes a saturation: for large average trustworthiness, node $i$ only receives a quantity $f h^* k_{i}$ that grows with the number of neighbours but not with the value of $\overline{h_{i}}$.

Finally, the third term accounts for the dependence of the current trustworthiness on its {\it speed of change}. In particular, $\delta h_i$ increases with the difference between the current and previous trustworthiness values, while the minimum operator implies that only negative recent changes are considered. Therefore, the coefficient $d$ tunes the amplification of negative events and introduces an asymmetry between positive and negative trustworthiness variations. In a sense, it measures the susceptibility of a population to panic. For simplicity, we shall refer to $d$ as ``panic factor''. The exact definition of $\delta h_i$ can be found in appendix \ref{sec:model_specs} or \cite{Batista2015}.

How real is our notion of perceived trustworthiness $h_i$? How could it be measured, for example? As mentioned above, one clear example are the CDSs of companies, which directly price the default 
probability as seen by market participants. Another possibility is to gauge the trustworthiness of individuals and firms through surveys, as discussed in \cite{Glaeser2000}, echoing a  
concern expressed by Putnam \cite{Putnam1995}: {\it since trust is so central to the theory of social capital, it would be desirable to have strong behavioural
indicators of trends in social trust or misanthropy. I have discovered no such behavioural measures.} Even if there is still a lot to be done in order to devise faithful, quantitative indicators of 
trustworthiness in general, it is highly plausible that the final answer will not be a single real variable as we assume, but a more complex, higher dimensional object. Nevertheless, we 
believe that the results obtained below, in particular those pertaining to the co-existence of different equilibria where collective trust is present or absent, will survive in more elaborate 
models of trustworthiness.

\subsection{Network dynamics}

We now specify how links in the network are created or broken depending on the trustworthiness of the nodes. Since the latter depends itself on the degree of the nodes and on its dynamics, we end up with a model of coupled trustworthiness/network dynamics which shows interesting properties, much as in \cite{Ehrhardt2006}.

At each time step, we choose a pair of nodes at random, say $(i,j)$, characterized by their trustworthiness $h_{i}$ and $h_{j}$. The total number of nodes is constant in time and equal to $N$. The global average (over all nodes) of $h_i$, which characterizes the overall confidence level in the network, is denoted by $\overline{h}=\sum_{i} h_i/N$.

\subsubsection{Link creation}

If there are no links between $i$ and $j$, the probability $\Pi_{ij}^+$ that they decide to do business together is
\begin{equation}
	\Pi_{ij}^+ = \frac{r}{N} \frac{z_{ij}}{1 + z_{ij}},
\end{equation}
where $0 < r < N$ is the a priori propensity to enter into a business relation (the factor $1/N$ is discussed below) and $z_{ij} \geq 0$ is a modulating factor that depends on the trustworthiness $h_{i}$ and $h_{j}$ as follows:
\begin{equation}\label{z:def}
	z_{ij} = e^{\alpha \overline{h} - \beta |h_{i}-h_{j}|},
\end{equation}
where $\alpha, \beta$ are two positive parameters. Therefore, a small value of $z$ implies a small probability of link formation. The term $\alpha \overline{h}$ attempts to capture the idea that a trustful society eases the creation of new collaborations or business relations, i.e. that {\it a rising tide lifts all boats}. This is the essential virtue of trust that we discussed in the introduction: it acts as a catalyst to exchange and activity, an effect that we attempt to model through $\alpha$. It is quite clear that together with Eq. (\ref{h:def}) above, this term can lead to a virtuous circle -- more confidence leads to a more connected society which in turn leads to more confidence.

The second term $-\beta |h_{i}-h_{j}|$ decreases $z$ and is consequently detrimental to link creation. This attempts to account for ``homophily'', i.e. the intuitive fact that two entities with very similar credit level are more likely to conduct business together than less comparable peers \cite{Ehrhardt2009,McPherson2001a,Marsden1988,Coleman1958,Pin2008}.

Instead of coupling $z_{ij}$ to the overall confidence level $\overline{h}$, one could have imagined to use only the ``local'' trustworthiness
$h_i + h_j$. We have in fact investigated a generalized model in which
\begin{equation}
	z_{ij} = e^{\alpha \overline{h} + \alpha' (h_i + h_j - 2\overline{h}) - \beta |h_{i}-h_{j}|},
\end{equation}
where the $\alpha'$ term captures deviations from the global average. We have found numerically that the new $\alpha'$ term does not change much the phenomenology of the model. This will be confirmed by the mean-field approximation below. We will thus set henceforth $\alpha'=0$.

\subsubsection{Link destruction}

If there is a link between the chosen pair $(i,j)$, it is destroyed with probability
\begin{equation}
	\Pi_{ij}^- =  \frac{1}{1 + z_{ij}} \, \in [0,1],
\end{equation}
which tends to unity when $z \ll 1$, i.e. when average confidence is very negative, or when homophily is strong ($\beta \gg 1$), both being detrimental to maintaining relationships. The specific choice for $\Pi_{ij}^{\pm}$, and the factor $N^{-1}$ in front of $\Pi_{ij}^+$, can be understood by calculating the probability $P_{ij}$ that the link between $i$ and $j$ exists in the stationary state. Assuming $z_{ij}$ to be time independent, $P_{ij}$ is the solution of
\begin{equation}
	\Pi_{ij}^+ (1-P_{ij}) - \Pi_{ij}^- P_{ij} = 0 \implies P_{ij} = \frac{rz_{ij}}{rz_{ij}+N} \underset{r \ll N}{\approx} \frac{rz_{ij}}{N}.
\end{equation}
Therefore, when $r, z$ are both of order unity, the probability that a link exists is of order $1/N$ and the typical degree of a node is itself of order $zr = O(1)$. This is the scaling needed in order to have a non trivial dynamics in the limit $N \to \infty$.

\section{Numerical results}

We have numerically investigated this model in detail for various values of its six parameters: $f,h^*,d$ for  trustworthiness and $\alpha, \beta, r$ for link creation/destruction. Some initial conditions for the $h$'s and for the state of the network also need to be specified to run the dynamics. It turns out that as soon as $N$ is somewhat large (i.e. $N\gtrapprox50$), and for some regions in parameter space, the dynamics of the model becomes {\it history dependent}, in the sense that starting from an empty network (no links at all) or a full network (all links are present) leads to completely different stationary states -- at least over time scales that can be reached in simulations and hence in reality as well (if our model captures anything of reality).

\begin{figure}[!htb]
	\subfloat[Initial conditions: $\tilde{L}(t=0)=0$. When $d=0$, the network is mostly dense for $\beta\lessapprox1.50$. If $d$ increases, the maximum $\beta$ which allows for the dense state decreases. When $d\gtrapprox2$, crash phenomena start to take place and $\tilde{L}$ in the dense state is lower than before. As $d\rightarrow4$, the dense state eventually becomes unreachable.\label{fig:hmapj00}]{%
		\includegraphics[scale=0.45]{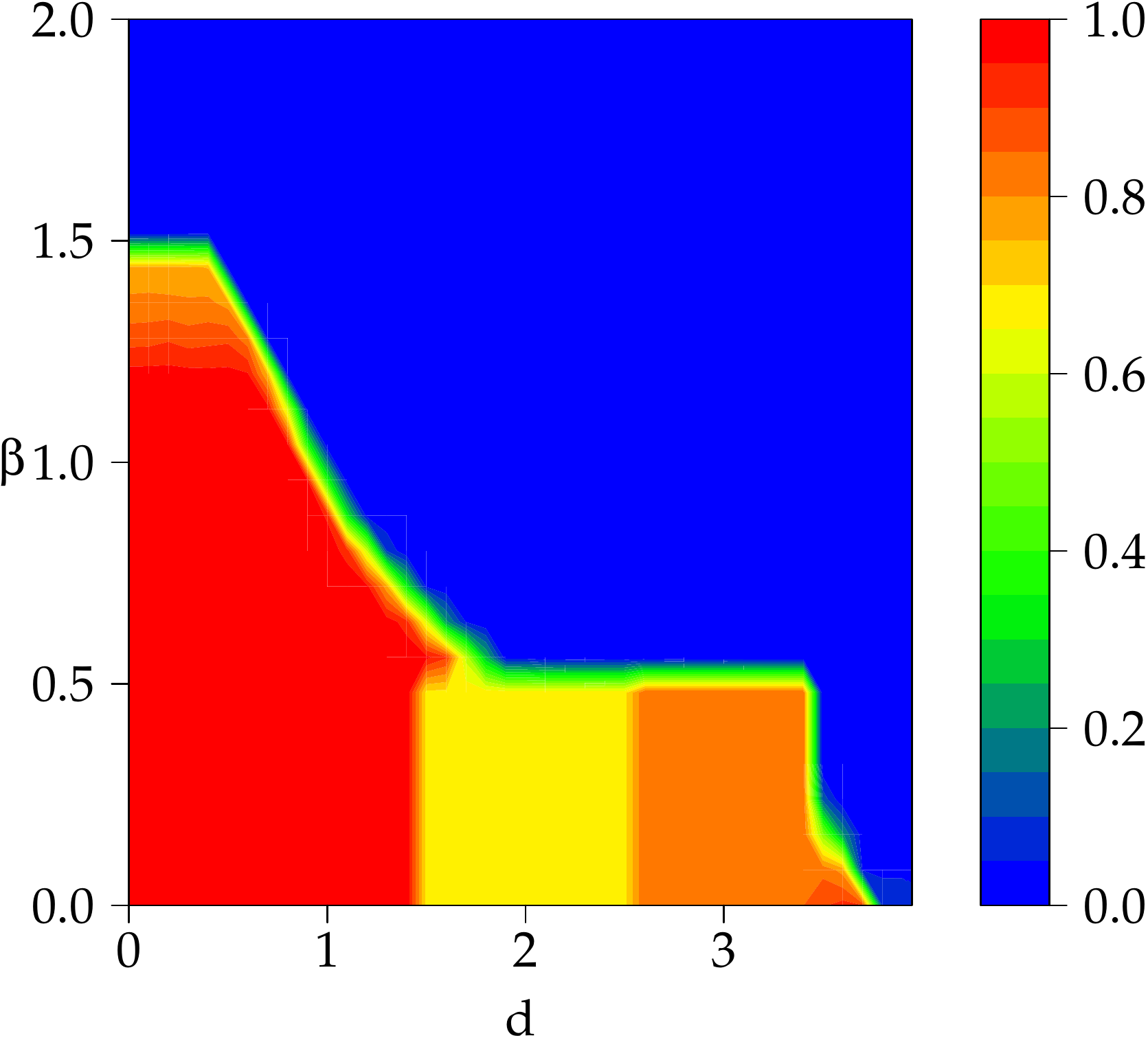}%
		}\hfill
	\subfloat[Initial conditions: $\tilde{L}(t=0)=0.9$. The region of the parameter space where the dense state is not the preferred stationary state is smaller than in Fig. \ref{fig:hmapj00} and the absolute sparse state ($\tilde{L}\approx0$) is not clearly visible. Further numerical calculations indicate that we would observe the absolute sparse state with these initial conditions beyond $\beta\approx10$.\label{fig:hmapj09}]{%
		\includegraphics[scale=0.45]{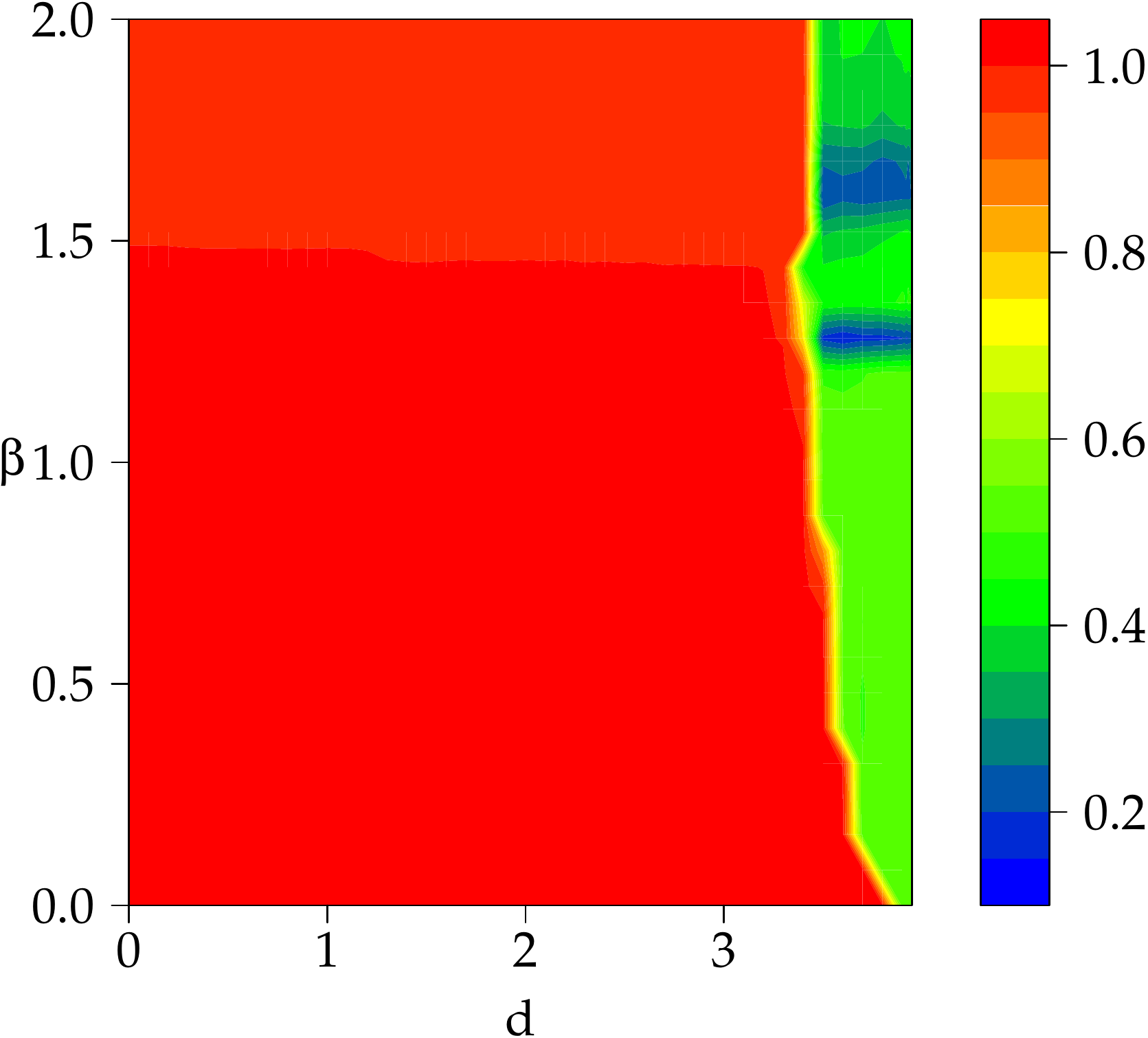}%
	}
	\caption[Content]{Average density of links $\tilde{L}$ for varying $d$ and $\beta$ and for two different initial conditions (sparse and dense). $N=200$, $\alpha=1$, $f=1$, $h^*=5$, $r=1$, $100$ runs and after $10^{5}$ time steps. Regardless of the initial condition, there are two distinct regions in the parameter space, which correspond to two different stationary states, with a sharp transition in between. The red area in the plot corresponds to a dense network ($\tilde{L}\approx1$) and the blue area corresponds to a sparse network ($\tilde{L}\approx0$).}
	\label{fig:hmap}
\end{figure}

\begin{figure}[!htb]
	\subfloat[$\tilde{L}$ \label{fig:hyst_l}]{%
		\includegraphics[scale=0.45]{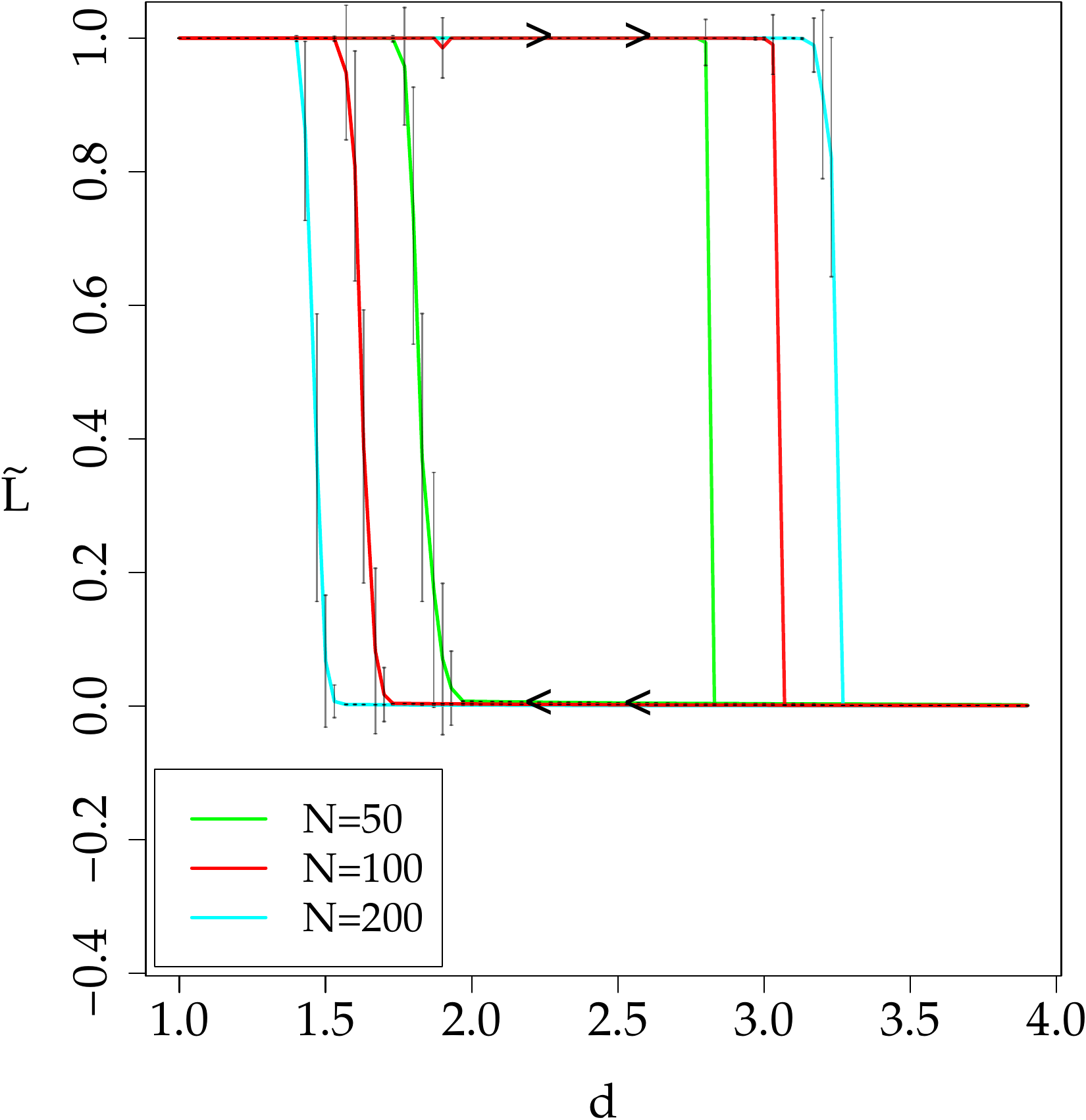}%
		}\hfill
	\subfloat[$\tilde{h}=\frac{\overline{h}}{fh^{*}N}$ \label{fig:hyst_h}]{%
		\includegraphics[scale=0.45]{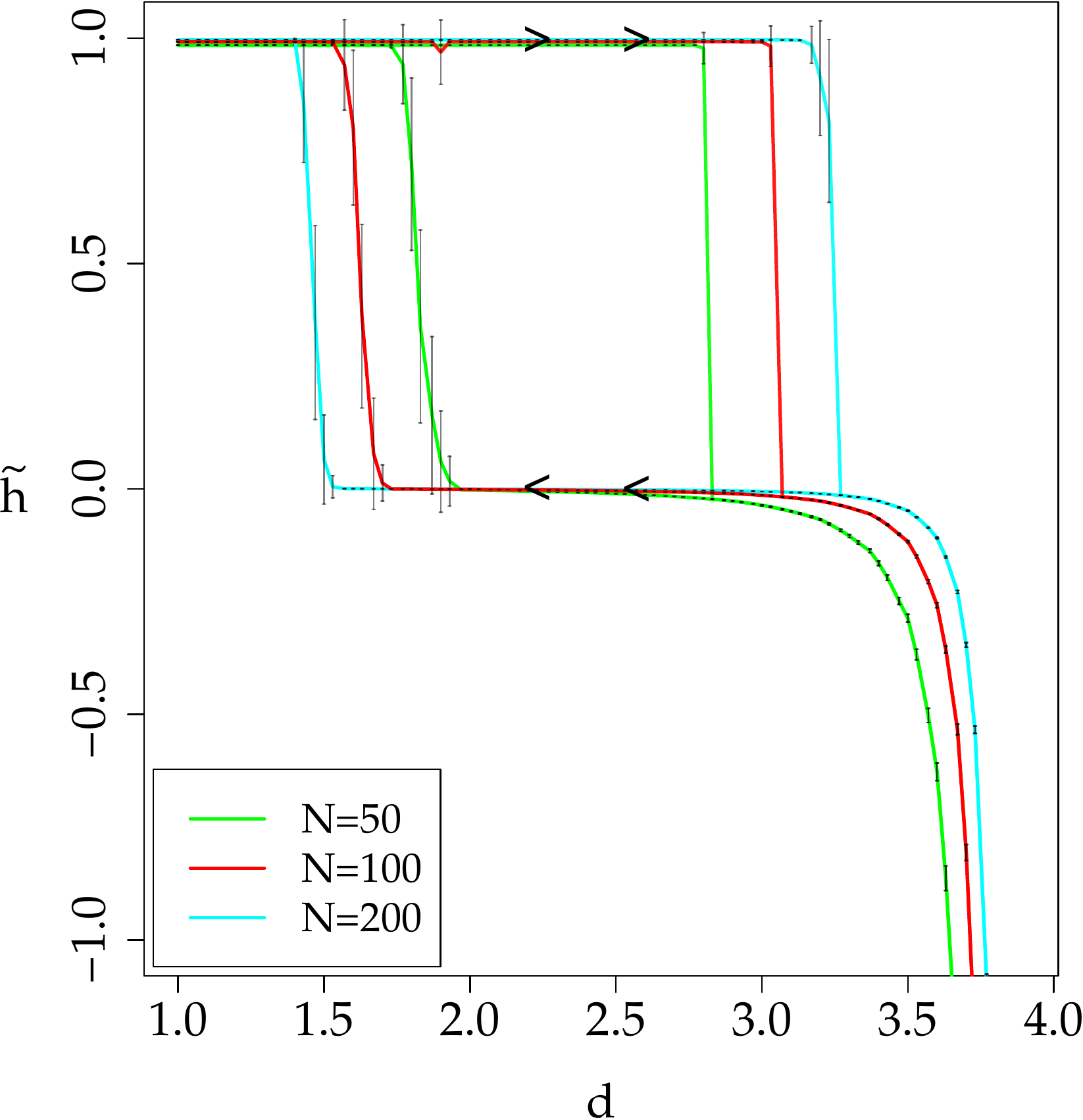}%
	}
	\caption[Content]{Path along $d=2\beta$ for $N=50,100,200$, $\alpha=1$, $f=1$, $h^*=5$, $r=1$, $100$ runs and $10^{6}$ time steps per point. The average density of links $\tilde{L}$ is shown on the left. The direction along the hysteresis path, in which $\tilde{L}$ ranges from $0$ to $1$, is represented by black arrows. The plot of the average trustworthiness $\overline{h}$ scaled by $fh^{*}N$, which we call $\tilde{h}$, is on the right. When $d$ is small and the number of links of the network approaches $\frac{1}{2}N(N-1)$, i.e. $\tilde{L}\rightarrow 1$, $\tilde{h}=\frac{\bar{h}}{fh^{*}N} \rightarrow 1$. When $d\rightarrow4$, $\tilde{h}\rightarrow-\infty$ and $\tilde{L}\rightarrow0$. The change from one state to the other occurs discontinuously, as observed in first order phase transitions.}
	\label{fig:hyst}
\end{figure}

The most important parameters of our model appear to be the homophily parameter $\beta$ and the panic factor $d$. This will be justified within our mean-field approximation below: as long as the confidence parameter $\alpha$ is not vanishingly small and $r$ is large enough, the phenomenology of the model is mostly determined by $\beta$ and $d$. We have therefore plotted the phase diagram of the model in the $(d,\beta)$ plane, and the results are shown in Fig. \ref{fig:hmap}. We represent the average density of links $\tilde{L} = \langle k \rangle / N$  of the network in a color code, starting from an empty network at $t=0$ (Fig. \ref{fig:hmapj00}) or from a densely connected network (Fig. \ref{fig:hmapj09}). Similar patterns appear when one represents the average confidence $\overline{h}$ instead. One observes a clear boundary line $\beta_c(d)$ separating two distinct phases: one in which the network is sparse in the stationary state, corresponding to a low average confidence $\overline{h}$, and another in which the network is dense, corresponding to a high average confidence $\overline{h}$. However, this boundary line shifts to significantly higher values when one starts from an already dense network. In other words, there is a large crescent region in phase space where the two outcomes (sparse or dense) are possible, and where the initial condition determines the fate of the network. Another way to illustrate this is to show the evolution of the density of links and of the average trustworthiness $\overline{h}$ as a function of $d$ as one cycles along the line $d = 2\beta$ as in Fig. \ref{fig:hyst_l} and Fig. \ref{fig:hyst_h}.

\begin{figure}[!htb]
	\subfloat[$\tilde{L}$ \label{fig:hyst_h0_l}]{%
		\includegraphics[scale=0.45]{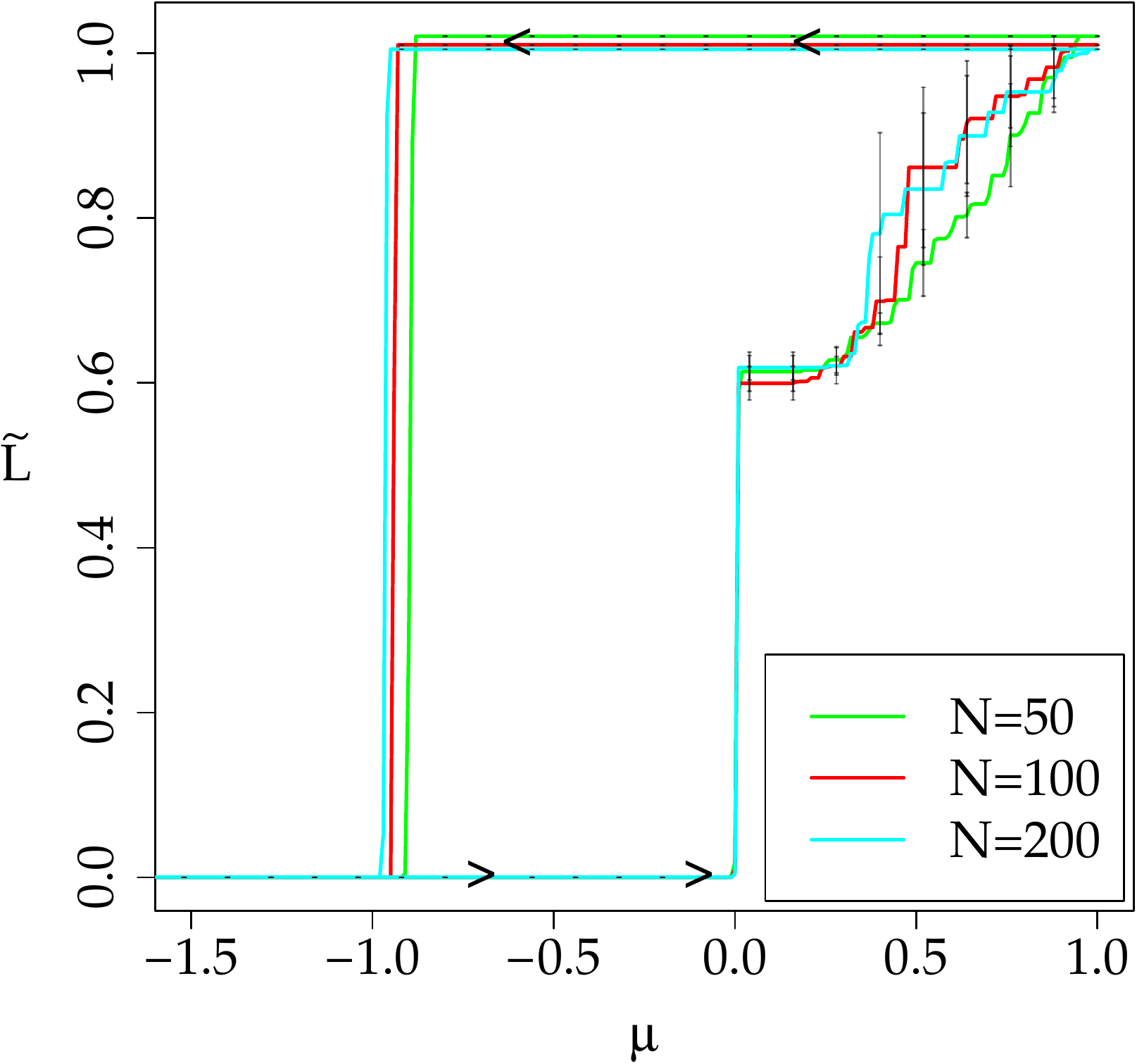}%
		}\hfill
	\subfloat[$\tilde{h}=\frac{\overline{h}}{fh^{*}N}$ \label{fig:hyst_h0_h}]{%
		\includegraphics[scale=0.45]{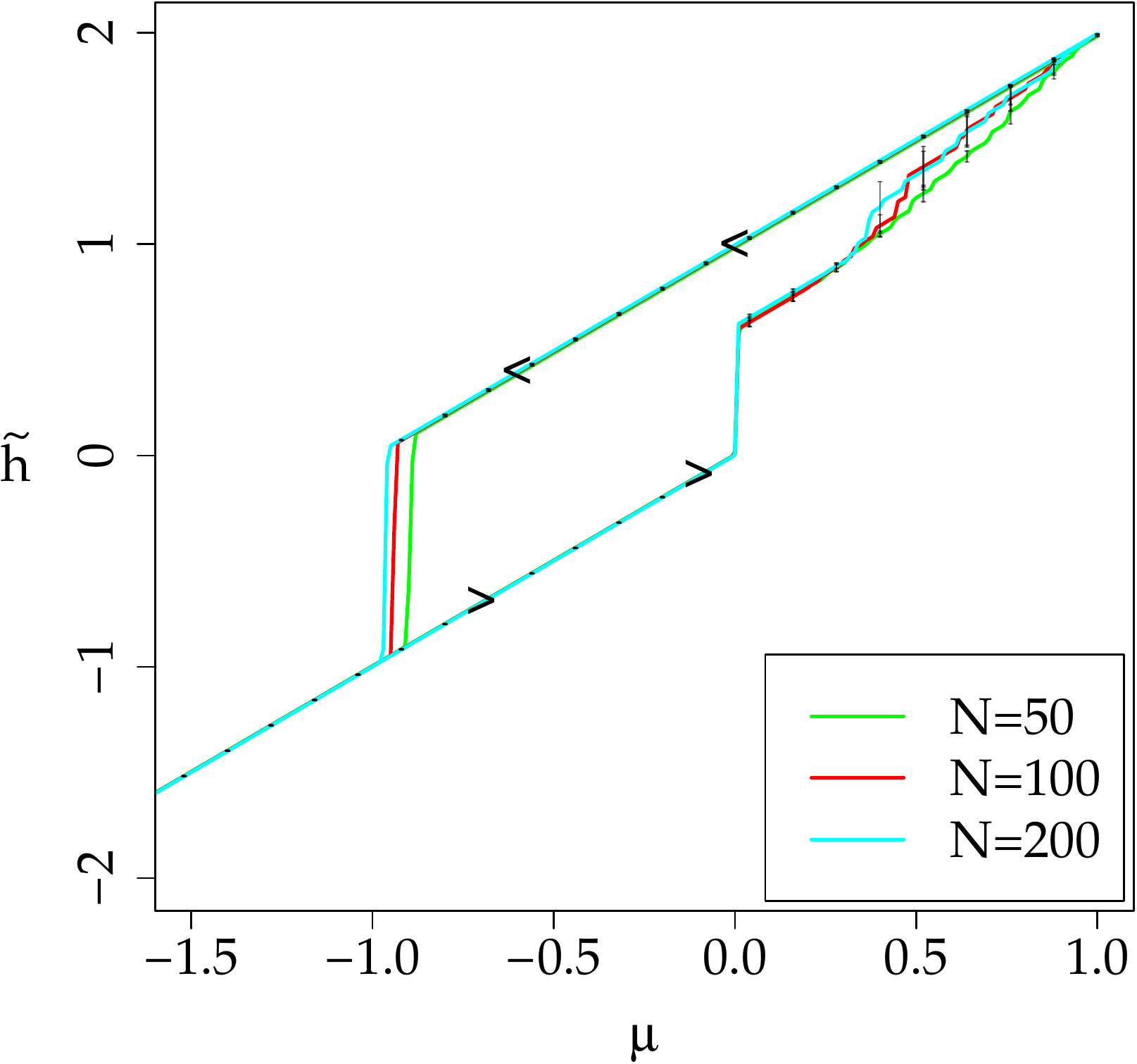}
	}
	\caption[Content]{Path along $\mu=m \cdot fh^{*}N$ for $N=50,100,200$, $\alpha=1$, $f=1$, $h^*=5$, $r=1$, $d=2$, $\beta=2$, $100$ runs and $10^{6}$ time steps per point. $m$ is the (time-dependent) common shift added to the original intrinsic trustworthiness of each node $h_{i,0}$. The direction along the hysteresis path, in which $\tilde{L}$ ranges from $0$ to $1$, is represented by black arrows. When we start at $\mu=1$, the network is dense and $\tilde{L}\approx1$. If we continuously decrease $\mu$ the network disintegrates ($\tilde{L}\approx0$) when $\mu \lessapprox -1$. Then, if we increase $\mu$ the network will switch back to the dense state ($\tilde{L}\approx1$) only when $\mu \gtrapprox 0$. The coexistence of two different equilibria allows the system to be trapped in one of these states even if the other is more favourable. Besides, we observe discontinuities when the system jumps from one state to the other.}
	\label{fig:hyst_h0}
\end{figure}

\begin{figure}[!htb]
	\centering
	\includegraphics[scale=0.45]{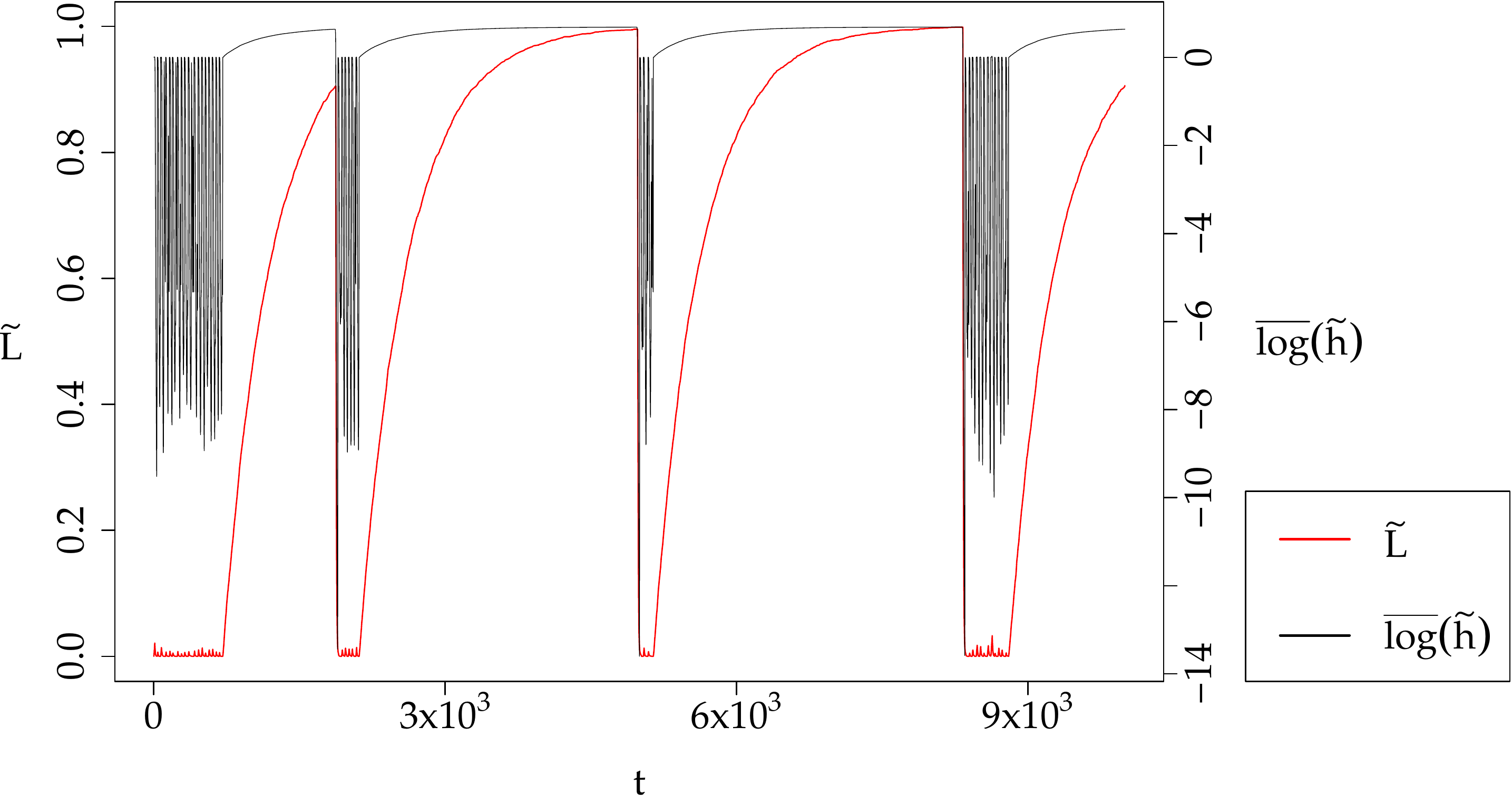}
	\caption[Content]{$\tilde{h}$ and $\tilde{L}$ (single run) for $N=100$, $\alpha=1$, $f=1$, $h^*=5$, $d=3.9$, $\beta=0$ and $r=1$. $\tilde{L}(t=0)=0$ and the function $\overline{ \log }$ is defined as $\overline{ \log }(x)= \mathrm{sign} (x).\overline{ \log }(|x|+1)$. We confirm that the system keeps switching between two distinct equilibria: one consisting in a dense network ($\tilde{L}\approx1$) and the other in a sparse network ($\tilde{L}\approx0$). The dense state is typically characterized by positive average trustworthiness $\tilde{h}$ which grows steadily in time towards the asymptotic value. On the other hand, the average trustworthiness $\tilde{h}$ in the sparse state tends to be negative and oscillates wildly. The transitions from the sparse state to the dense state are smooth and steady, while the transitions from the dense state to the sparse state, which are triggered by random fluctuations that break links and cause cascade phenomena, are quick and abrupt.}
	\label{fig:srun_crashes}
\end{figure}

For small $N$ (but still large enough to be of practical interest, say $N \lesssim 100$) the system can in fact alternate between these two states,  leading to interesting endogenous crises -- i.e. large swings between high confidence and low confidence that are not due to any particular event, but are the result of the noisy evolution of a system for which two very different equilibrium states coexist -- see Fig. \ref{fig:srun_crashes}. As $N$ grows larger and larger, the probability to jump from one state to another becomes exponentially small, a typical behaviour of physical systems undergoing a first order phase transition (see below for a discussion of this point within a mean-field approximation). However, interesting dynamics will follow from the time-variation of parameters. A suggestive numerical experiment is to let the average value $m$ of the intrinsic trustworthiness $h_{i,0}$ slowly evolve with time, in order to model a progressive shift of the objective state of the economy. When the system is in the coexistence region, one observes a succession of booms and crises, corresponding to jumps between the two underlying equilibrium states -- see Fig. \ref{fig:hyst_h0_l} and Fig. \ref{fig:hyst_h0_h}.

An analytic description of the dynamics of crisis and recovery can be performed, in particular when $\beta = 0$ and close to the complete instability limit $d=4$, which is derived in appendix \ref{sec:d_bounds}.  The interested reader is referred to \cite{Batista2015} for further details. We now turn to a mean-field approximation that accounts relatively well for our numerical observations.

\section{A mean-field analysis}

\subsection{Warm-up: Erd{\"o}s-R{\'e}nyi}

Let us start by adopting a kinetic view of the standard Erd{\"o}s-R{\'e}nyi network with $N$ nodes. At each time step $t$, a link is randomly chosen among the $\frac{N(N-1)}{2}\approx\frac{N^{2}}{2}$ possible links. Following the same notation as before, the probability to create a link is $\Pi^{+}=\frac{r}{N}\frac{z}{1+z}$, where, for the time being, $r$ and $z$ are constants. If the link is already present, the probability that it is destroyed is $\Pi^{-}=\frac{1}{1+z}$. We introduce the time-dependent degree distribution $P(k,t)$, i.e. the probability that a randomly chosen node has exactly $k$ outgoing links at time $t$. The probability that this node changes from $k\rightarrow k+1$ in the next time step $t+1$ is
\begin{equation}
	W^{+}(k)=\frac{2}{N^{2}}(N-k)\frac{rz}{N(1+z)},
\end{equation}
while the probability to change from $k\rightarrow k-1$ in the next time step is
\begin{equation}
	W^{-}(k)=\frac{2}{N^{2}}\frac{k}{(1+z)}.
\end{equation}
Making time a continuous variable leads to the following Master equation for $P(k,t)$:
\begin{equation}
	\frac{\partial P(k,t)}{\partial t} = \frac{2}{N^{2}(1+z)}\left[rz\frac{N-k+1}{N}P(k-1,t)+(k+1)P(k+1,t)-\left(rz \frac{N-k}{N}+k\right)P(k,t)\right].\label{eq:mastereq_erdos}
\end{equation}
By inspection, one finds that $P_{0}(k)=C_{N}^{k}q^{k}(1-q)^{N-k}$ is a stationary solution of Eq. (\ref{eq:mastereq_erdos}), as it should be, provided
\begin{equation}
	q=\frac{zr}{zr+N}.
\end{equation}
The average degree $\langle k\rangle$ and the corresponding variance are then given by:
\begin{eqnarray}
	\langle k\rangle & = & Nq=\frac{Nzr}{zr+N} \underset{{N \to \infty}} \approx zr \label{eq:erdos_k_mean}\\
	\langle k^{2}\rangle-\langle k\rangle^{2} & = & Nq(1-q)\label{eq:erdos_k_var}
\end{eqnarray}

The following sections extend the above calculation to the case where $z$ self-consistently depends on the trustworthiness of the nodes.

\subsection{Coupling with the average trustworthiness $\overline{h}$}

We now consider the baseline case where $z=e^{\alpha \overline{h}}$, with $\alpha > 0$ and $\overline{h}$ the average trustworthiness of the population. For the time  being, we discard all homophily effects or feedback loops (i.e. $\beta=d=0$).

We first assume that the average intrinsic trustworthiness $h_{i,0}$ has a zero mean, $m=0$. This is an interesting situation since it does not break the $h \to -h$ symmetry, i.e. collective trust or distrust are a priori equally probable outcomes. Averaging Eq. (\ref{h:def}) over all nodes and using a mean field argument, i.e neglecting all fluctuations making all $h_i$ different, we find
\begin{equation}
	\overline{h}=f h^* \langle k\rangle \tanh\left(\frac{\overline{h}}{h^*}\right).\label{eq:mf_hbar_selfcons}
\end{equation}
This approximation is certainly justified in the dense limit $\langle k\rangle \gg 1$, but breaks down for small $\langle k\rangle$, in particular when $\langle k\rangle < 1$. In this latter case the network does not percolate and, in the absence of a giant component, no collective behaviour is possible. In this case, the only solution to Eq. (\ref{eq:mf_hbar_selfcons}) is $\overline{h}\approx0$.

Suppose for simplicity that $f \langle k\rangle$ is somewhat larger than unity (say $5$ or more), then $|\tanh(\frac{\overline{h}}{h^{*}})|\approx 1$ and Eq. (\ref{eq:mf_hbar_selfcons}) has two possible solutions: $\overline{h} \approx \pm fh^* \langle k\rangle$ \footnote{For \unexpanded{$f \langle k\rangle >1$} but not so large, the qualitative discussion below remains valid, up to prefactors of order unity.}. Now we can plug these solutions in Eq. (\ref{eq:erdos_k_mean}), which yields a second self-consistent equation:
\begin{equation}
	\langle k\rangle=\frac{Nr}{r+N e^{\mp \varphi \langle k\rangle}}, \qquad \text{where } \varphi \equiv \alpha f h^{*}. \label{eq:mf_hbar_k_mean}
\end{equation}

\subsubsection{The positive trust self-consistent solutions}

Let us focus first on the case where a {\it positive average} trustworthiness appears, corresponding to the minus sign in the exponential in Eq. (\ref{eq:mf_hbar_k_mean}). Assume first that $\varphi \langle k\rangle \gg \log N$. Then, the second term in the denominator is completely negligible and $\langle k\rangle \approx N$, which obeys the above hypothesis provided $\varphi \equiv \alpha f h^* > \log N/N$, which we will assume in the following. This corresponds to a self-sustained ``euphoric state'' where the network is full and confidence at its  peak. This solution always exists unless $\varphi$ is vanishingly small:  in the absence of the detrimental effects studied below, a dense network should appear due to the positive feedback term that favours link formation when confidence rises.

A second, sparse but percolating (i.e. with a giant component) solution can also exist. To see that this is the case, assume now that $z=\mathcal{O}(1)$. Then, Eq. (\ref{eq:mf_hbar_k_mean}) leads to $\langle k\rangle = zr$, where
\begin{equation}
	z = e^{\varphi z r}.
\end{equation}
This self-consistent equation depends on the product $\varphi r$:
\begin{itemize}
	\item When $\varphi r > e = 2.71..$, there is no solution to this equation. Only the dense network solution described above
	exists.
	\item When $\varphi r < e = 2.71..$, on the other hand, there are 2 solutions $z_<$ and $z_>$, one stable corresponding to a sparse, but trustful network, and a dynamically unstable one, which is nevertheless interesting since the associated value for $\langle k\rangle^* = z_> r$ is the critical value above which a sparse network is unstable and flows towards the fully connected solution above. Said differently, if the spontaneous fluctuations around the stable solution $\langle k\rangle = z_< r$ are not strong enough to reach $\langle k\rangle^*$ with appreciable probability, the sparse network will appear dynamically stable. This is indeed the case when $\varphi$ is small enough.
\end{itemize}

\subsubsection{A negative trust self-consistent solution}

An important question at this point is whether this model also allows for the existence of sustained negative average trustworthiness values $\overline{h}<0$, i.e. a connected, but suspicious society. This would correspond to the positive sign in the exponential in Eq. (\ref{eq:mf_hbar_k_mean}). In this case, the solution for large $N$ is:
\begin{equation}
	ye^{\varphi y}=r, \qquad y=\langle k\rangle.\label{eq:mf_nhbar_k_mean_1}
\end{equation}
When $\varphi r \ll 1$, the solution of Eq. (\ref{eq:mf_nhbar_k_mean_1}) is $\langle k\rangle\approx r$, therefore when $r>1$ the solution with negative $\overline{h}$ is indeed self-consistent. Hence a self-sustained state of distrust in a sparse network (but with a giant component) is possible when a) $\varphi$ is small enough (i.e. distrust is not too detrimental to link formation) and b) $r$ sufficiently large (i.e. agents meet often enough so that links are created even if the two parties are mutually suspicious). This corresponds, pictorially, to a ``wary'' society in which distrustful relationships are the norm.

On the other hand, if $\varphi r \gg 1$, we have
\begin{equation}
	\langle k\rangle\approx\frac{1}{\varphi}\left[\log(\varphi r)+\mathcal{O}(\log \log (\varphi r))\right].\label{eq:mf_nhbar_k_mean_2}
\end{equation}
Equation (\ref{eq:mf_nhbar_k_mean_2}) shows that as $\varphi$ grows, $\langle k\rangle$ decreases until the giant component disappears (when $\langle k\rangle < 1$) and the solution with $\overline{h}<0$ is no longer viable. For large $r$, this occurs for a certain value $\varphi_c \sim \log r + O(\log \log r)$. We have checked numerically that this ``wary society'' phase indeed exists in our model and is not an artifact of the mean-field approximation.

\subsubsection{Summary}

Summarizing, for $\varphi=\mathcal{O}(1)$ and $r>1$ there are {\it three} viable solutions, one corresponding to very dense networks and positive self-sustained collective trust, and the two other to sparse networks (but still percolating, $\langle k\rangle>1$), one with positive and one with negative self-sustained trust. These latter two solutions however disappear as $\varphi$ increases, beyond $\sim e/r$ for the former and $\sim \log r$ for the latter.

The above analysis assumed that the average intrinsic trustworthiness is $m=0$. When $m > 0$, the self consistent equation becomes:
\begin{equation}
	\overline{h}= m+f h^* \langle k\rangle \tanh\left(\frac{\overline{h}}{h^*}\right).\label{eq:mf_hbar_selfcons_m}
\end{equation}
Clearly, this equation now selects the dense, positive confidence solution as soon as $\alpha m$ is not vanishingly small. This is the situation we have considered in simulations.

\subsection{Coupling with speed of trust degradation}

We now study the influence of the panic parameter $d$ on the trustworthiness in Eq. (\ref{h:def}), i.e. the positive feedback effect that may trigger a link breaking avalanche when an increase of perceived risk takes place. We set the homophily term $\beta$ to zero for the time being and look into the general case in the next section.

As a warm-up exercise, let us compute the evolution of $\langle k\rangle_{t}=\sum_{k} k P(k,t)$ from Eq. (\ref{eq:mastereq_erdos}). Multiplying by $k$ and summing over $k$ yields
\begin{equation}
	\frac{d}{dt}\langle k\rangle_{t}=\frac{2}{(1+z)N^{2}}\left[zrN-(zr+1)\langle k\rangle_{t}\right].
\end{equation}
At equilibrium, with $\frac{d}{dt}\langle k\rangle_{t}=0$, we trivially recover the result in Eq. (\ref{eq:erdos_k_mean}):
\[
	\langle k\rangle_{eq}=\frac{Nzr}{N+zr}.
\]
For small deviations from equilibrium, $\langle k\rangle_{t}$ is described by an Ornstein-Uhlenbeck process that can be fully characterized from the knowledge of the variance of $k$.

Now, in our model with feedback we assume that all events contributing to lowering the degree of the nodes will lead to a decrease of trustworthiness. Restricted to events lowering the degree, this contribution can be written as
\begin{equation}
	\frac{d}{dt}\langle(\Delta k)_{-}\rangle=\frac{2}{(1+z)N^{2}}\Sigma_{k}kP(k,t)=\frac{2\langle k\rangle_{t}}{(1+z)N^{2}}.
\end{equation}
After $T=\frac{N^{2}}{2}$ time steps, which is the average time it takes to attempt to change the status of each link once, the total contribution to degree decrease is
\begin{equation}
	\langle(\Delta k)_{-}\rangle \approx \frac{\langle k\rangle}{1+z}.
\end{equation}
Again in a mean-field spirit, the resulting expression for $z$ is
\begin{equation}
	z=e^{\alpha \overline{h}-d\langle(\Delta k)_{-}\rangle},
\end{equation}
meaning that the stronger the activity that decreases connectivity, the smaller the value of $z$ and hence the larger the probability of breaking further links. There is also a second contribution to $\min(0, \delta h)$ arising from the time fluctuations of $\overline{h}$ itself, but it is much smaller in the equilibrium region we are focusing on.

Hence, we find a set of self-consistent equations valid when $f \langle k\rangle \gg 1$ and $\overline{h}>0$:
\begin{eqnarray}
	z &=& e^{\langle k\rangle\left(\varphi -2d\frac{1}{1+z}\right)}\label{eq:mf_wpanic_selfcons_1}\\
	\langle k\rangle &=& \frac{Nzr}{N+zr}\label{eq:mf_wpanic_selfcons_2}.
\end{eqnarray}

Let us study the possible solutions to Eq. (\ref{eq:mf_wpanic_selfcons_1}) and Eq. (\ref{eq:mf_wpanic_selfcons_2}). Suppose first that $N\ll rz$. In this case, we have from Eq. (\ref{eq:mf_wpanic_selfcons_2}) that $\langle k\rangle\approx N$. The self-consistent Eq. (\ref{eq:mf_wpanic_selfcons_1}) then leads to
\[
	z \approx e^{\varphi N},
\]
which is indeed such that $N\ll rz$ provided that $\varphi \gg\frac{\log (N)}{N}$. This solution corresponds to such dense a network that the downwards degree fluctuations cannot destabilize it, at least locally.

\begin{figure}[!htb]
	\centering
	\includegraphics[scale=0.45]{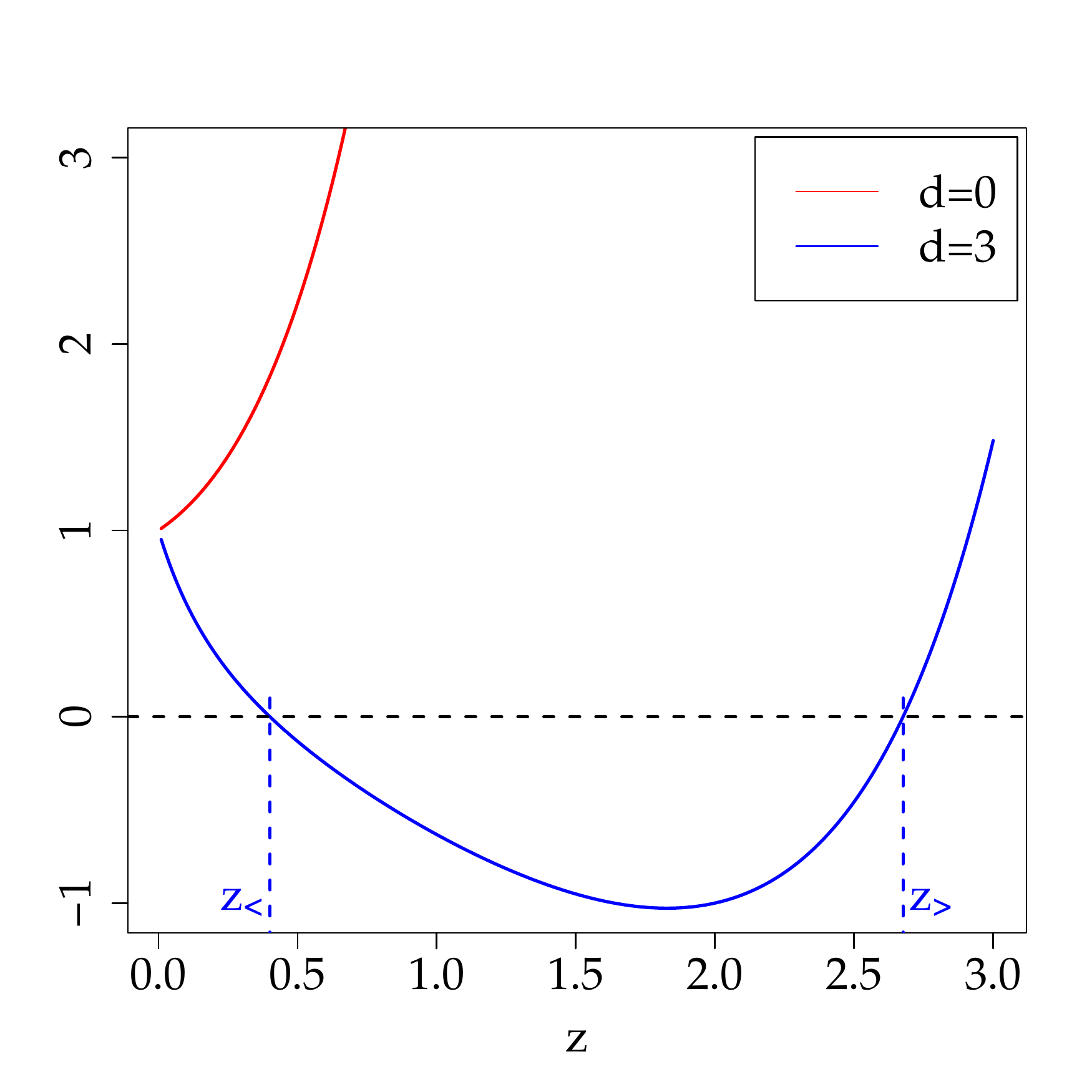}
	\caption[Content]{Graphical representation of Eq. (\ref{eq:mf_wpanic_selfcons_3}) for $d=0$ and $d=3$ and $\varphi=2, r=1$. For $d=0$ (red line) the function $f(z) := e^{\varphi r z-2d\frac{r z}{1+z}} - z$ has no zeros, which means that $z=\mathcal{O}(1)$ and $\langle k \rangle \approx rz$ are not compatible conditions in this case. However, for large enough $d$, for example $d=3$ (blue line), two solutions emerge: $z_{<}<1$ and $z_{>}>1$. The former is dynamically stable, while the latter is unstable and characterizes the critical transition path from the dense solution towards the sparse solution.}
	\label{fig:mf_diagram}
\end{figure}

However, there might coexist a second solution, even for values of $\varphi$ where it would not exist for $d=0$. Suppose now  that $z=\mathcal{O}(1)$ and $\langle k\rangle\approx rz$, which we assume to be larger than $1$ to allow for non-zero collective trust $\overline{h}>0$ to exist and be locally stable. The self-consistent equation now reads
\begin{equation}
	z=e^{\varphi r z-2d\frac{r z}{1+z}}.\label{eq:mf_wpanic_selfcons_3}
\end{equation}
It is clear that there is no solution to Eq. (\ref{eq:mf_wpanic_selfcons_3}) when $d$ is small and $\varphi r > e$. However, there is a critical value of $d$, denoted by $d^*$, above which Eq. (\ref{eq:mf_wpanic_selfcons_3}) has two solutions: $z_< <1$, which is stable at least for $d$ not too large, and $z_> > 1$, which is unstable. This is illustrated in Fig. \ref{fig:mf_diagram}. As $d$ increases further, $z_<$ becomes smaller and smaller and at one point becomes itself unstable, leading to limit cycle dynamics. This small $z_<$ solution however corresponds to a completely disconnected network.

The existence of a second, sparse solution for large enough $d$ corresponds well to our numerical observations: the network attempts to connect but trustworthiness is small and cannot grow because it is killed by spontaneous negative fluctuations.

The intermediate, unstable solution $z_>$ is also interesting as it again characterizes the critical transition path from the dense solution towards the sparse solution (and vice versa). For large $d$, one finds $z_> \approx 2d/\varphi$, corresponding to a characteristic average degree $k_> \approx 2dr/\varphi$. When $k_>$ is much smaller than $N$, the dense solution has an exponentially small (in $N$) probability of spontaneous destabilisation. However, as $k_>$ increases towards $N$, fluctuation induced crash events become more and more frequent, as shown in Fig. \ref{fig:srun_crashes}.

\subsection{Homophily}

We finally turn to the influence of homophily, i.e. the $\beta$ term in the definition of $z$ in Eq. (\ref{z:def}). Here we assume, as in \cite{Ehrhardt2006}, that the network is at all times an Erd{\"o}s-R{\'e}nyi network with a time dependent density of links $q_t=\langle k \rangle_t/N$. We also assume, as above, that the network is well-formed, with $f \langle k \rangle_t$ somewhat larger than unity so that one can assume that for most nodes, the following approximation holds:
\begin{equation}
	h_i \approx h_{i,0} + f h^* k_i.
\end{equation}
Again, two cases should be considered. One corresponds to dense networks, such that $\langle k \rangle \sim N$. In this case, fluctuations of node degree are at most of order $\sqrt{N}$. In fact, the homophily term leads to cliques of connected nodes with a relatively homogeneous degree, so we expect these fluctuations to be much smaller than $\sqrt{N}$. Therefore, one can estimate $z$ as
\begin{equation}
	\log z \approx \alpha N - \beta \sqrt{N},
\end{equation}
which shows that unless $\alpha$ is very small, the highly connected phase is not destabilized by homophily.

In the case of sparse but percolating networks with $\langle k \rangle > 1$, the dispersion of trustworthiness that prevents links from forming has two distinct origins. One is the intrinsic heterogeneity of the nodes, measured by the root mean square $\sigma$ of the fields $h_{i,0}$. The second is the degree heterogeneity which, for an Erd{\"o}s-R{\'e}nyi network with $q_t=\mathcal{O}(N^{-1})$, is given by $\sqrt{Nq_t}=\sqrt{\langle k \rangle_t}$. Using $\langle k \rangle_t = zr$, valid in the sparse phase, one finally ends up with the following schematic estimate of the homophily term:
\begin{equation}
	\beta |h_i - h_j| \longrightarrow \beta \sqrt{c\sigma^2 + c'z r},
\end{equation}
where $c,c'$ are numerical constants of order unity. This leads to a new self-consistent equation for the link activity $z$ in the sparse phase:
\begin{equation}
	z = e^{\varphi r z - \beta \sqrt{c\sigma^2 + c' zr}}.
\end{equation}
It is graphically clear that this equation behaves much in the same way as Eq. (\ref{eq:mf_wpanic_selfcons_3}): for small $\beta$ and $\varphi r > e$, no solution exists except for dense networks. But as $\beta$ increases, two non-trivial solutions, $z_<$ and $z_>$ appear, corresponding to a sparse solution that is not able to connect because of the strong repulsion between different nodes.  This corresponds to the sparse phase observed in the phase diagram of the model for large $\beta$, see Fig. \ref{fig:hmap}.

\section{Conclusion\label{sec:conclusion}}

We have introduced, in the spirit of \cite{Marsili2004,Ehrhardt2006}, a highly stylized model for the asymmetric build-up and collapse of collective trust in a network where the links and the trustworthiness of the nodes dynamically co-evolve. The basic assumption of our model
is that whereas trustworthiness begets trustworthiness (meaning that a higher level of trustworthiness is more favourable to link formation), trustworthiness heterogeneities, both across nodes and in time, are detrimental to the network. In particular, panic also begets panic, in the sense that sudden drops of trust may lead to link breaking (or ``sell-offs'' in the context of financial markets) that further decreases trustworthiness. We have shown, using both numerical simulations and mean-field analytic arguments, that there are extended regions of parameter space where two equilibrium states coexist: one corresponds to a favourable, well connected network with a high level of confidence
prevails, and the second is an unfavourable, poorly connected and low-confidence state. In these coexistence regions, sudden spontaneous jumps between the two states can occur. These transitions are not induced by any major catastrophe that would replace a favourable equilibrium by an unfavourable one, but rather by random fluctuations that trigger the switch between two {\it already existing equilibria}. When the system becomes large, however, these jumps become less and less frequent, unless an external parameter is changed -- corresponding, for example, to a measure of the overall economic activity that sets the average trustworthiness level. For large systems, the phenomenon of spontaneous crises is replaced by the notion of strong {\it history dependence}: whether the system is found in one state or in the other essentially depends on initial conditions: ergodicity is dynamically broken.

Our stylized model only aims at this stage to provide a generic (but certainly oversimplified) conceptual framework to understand how financial markets, or the economy as a whole, can shift so rapidly from a relatively efficient state to chaos, when nothing ``material'' has changed at all, when {\it our minds are no less inventive, our goods and services no less needed than they were last week}, as noted by President Obama. Our model illustrates Keynes remark: {\it a conventional valuation which is established as the outcome of the mass psychology of a large number of ignorant individuals is liable to change violently as the result of a sudden fluctuation of opinion due to factors which do not really make much difference} \cite{Keynes2006}. A theoretical challenge is of course to take our framework seriously and think about how such a model could be calibrated against data, for example using interbank loan networks (see e.g. \cite{Battiston2012}), CDS data or survey results as in \cite{Glaeser2000}. An obvious goal would be to obtain early warning signals for potential trust collapse and crises \cite{Squartini2013} that could, in some cases, look like precursor avalanches or ``crackling noise'' (see \cite{Sethna2001}, and for a recent review on this theme, \cite{Bouchaud2013}).

\section{Acknowledgements}
This work was partially financed by the EU ``CRISIS'' project (grant number: FP7-ICT-2011-7-288501-CRISIS) and Funda\c{c}{\~a}o para a Ci\^{e}ncia e Tecnologia.

\appendix
\section{Model specifications\label{sec:model_specs}}

The adjacency matrix at time $t$ is denoted by $J_{ij,t}$, while the trustworthiness of node $i$ at time $t$ is given by $h_{i,t}$. $N$ is the total number of nodes in the network and $k_{i,t}$ is the degree of node $i$ at time $t$, i.e., $k_{i,t}=\sum_{j}J_{ij,t}$.

At each time step $t$, the links between nodes are updated first. Then, the new trustworthiness of each node is computed.

Therefore, the evolution of the system at each time step happens in two distinct steps as follows.
\begin{enumerate}
	\item Create, destroy, or leave $sN,\ s\in]0,1]$, links untouched:\\
	\begin{eqnarray}\label{eq:model_specs_l}
		P(J_{ij,t}=0|J_{ij,t-1}=1) & =: & \Pi^{+}_{ij}=\frac{1}{1+z_{ij,t-1}}\\
		P(J_{ij,t}=1|J_{ij,t-1}=0) & =: & \Pi^{-}_{ij}=\frac{r}{N}\frac{z_{ij,t-1}}{1+z_{ij,t-1}}\ ,
	\end{eqnarray}
	where
	\begin{equation}\label{eq:model_specs_z}
		z_{ij,t} = e^{\alpha\overline{h_{t}}+\alpha'(h_{i,t}+h_{j,t}-2\overline{h_{t}})-\beta|h_{i,t}-h_{j,t}|} \text{ and } r\in\mathbb{R}^{+}.
	\end{equation}
	\\
	\item Update the trustworthiness values $h_{i}$:\\
	\begin{equation}\label{eq:model_specs_h}
		h_{i,t}=h_{i,0}+fk_{i,t} \tanh \left(\frac{1}{ck_{i,t}}\widetilde{P_{i,t}}\right)+d\cdot \min \left(0,\ \delta h_{i,t}\right),
	\end{equation}
	where:

	\begin{itemize}
		\item[] $\delta h_{i,t}=h_{i,t-1}-h_{i,t-2}+f\left[k_{i,t} \tanh \left(\frac{1}{fk_{i,t}}\widetilde{P_{i,t}}\right)-k_{i,t-1} \tanh \left(\frac{1}{fk_{i,t-1}}P_{i,t-1}\right)\right]$,
		\item[] $\widetilde{P_{i,t}}=\sum_{j}J_{ij,t}h_{j,t-1}$,
		\item[] $P_{i,t}=\sum_{j}J_{ij,t}h_{j,t}$,
		\item[] and $f,d\in\mathbb{R}^{+}$.
	\end{itemize}
\end{enumerate}
Regarding the first step, it is worth remarking that $\lim_{N\rightarrow\infty}\Pi^{+}_{ij}\cdot N^{2}\propto N$, which implies that the number of new links per node remains finite even for large $N$. $z_{ij,t}$ is a measure of the propensity of nodes $i$ and $j$ to link or remain linked at time $t$, which we assume to increase with $\overline{h_{t}}$ and $h_{i,t}+h_{j,t}-2\overline{h_{t}}$. On the other hand, we consider that $z_{ij,t}$ is bigger if $|h_{i,t}-h_{j,t}|$ is smaller, i.e., that the likelihood of node $i$ linking with node $j$ increases with the similarity of their perceived trustworthiness in the community (homophily).

The term $\min \left(0,\ \delta h_{i,t}\right)$, with its intrinsic asymmetry, is a proxy for the panic sentiment mentioned in the main text. Besides, $\widetilde{P_{i,t}}$ is the tentative cumulative trustworthiness of the peers of node $i$ at time $t$, while $P_{i,t}$ is the actual value.

We can view the parameter $s$ as a mere refresh rate in the algorithm but we can also interpret it as a measure of overall communication intensity between nodes.

\section{Panic factor $d$ and stability\label{sec:d_bounds}}
Let us consider the case where node $i$ ends up without any links at time $t_L+1$. Moreover, let us assume that $z_{ij,t}$ is small enough for us to neglect new links involving node $i$ as per Eq. (\ref{eq:model_specs_l}).
For the sake of simplicity, let us define $\tau:=t-t_L$. In this notation, node $i$ has at least one link at $\tau=0$ and becomes disconnected from the rest of the network at $\tau=1$. Moreover, let us define $h_n:=h_{i,\tau+n}$ and $h_\mathrm{init}:=h_{i,0}$.

Then, we have from Eq. (\ref{eq:model_specs_h}) that
\begin{equation}\label{eq:panic_san_hseq}
	h_n=h_\mathrm{init}+d(h_{n-1}-h_{n-2}),\quad n\geq2.
\end{equation}

In this scenario, Eq. (\ref{eq:panic_san_hseq}) defines the fate of node $i$, as it determines whether its trustworthiness $h_{i,t}$ enters an infinite downfall or not.

Equation (\ref{eq:panic_san_hseq}) can be re-written as
\begin{equation}\label{eq:panic_san_hseqmat}
	\mu_n=\Delta\mu_{n-2}+\nu,
\end{equation}
where:
\begin{eqnarray}
	\Delta & = & d\left[\begin{array}{cc}
		d-1 & d \\
	1 & -1\end{array}\right]\\
	\mu_n & = & \left[\begin{array}{cc}
	h_n & h_{n-1}\end{array}\right]^T\\
	\nu_n & = & h_\mathrm{init}\left[\begin{array}{cc}
	d+1 & 1\end{array}\right]^T.
\end{eqnarray}

After some computations, Eq. (\ref{eq:panic_san_hseqmat}) becomes

\begin{eqnarray}
	\mu_{2n+1} & = & \Delta^n (\mu_1-v)+v  \label{eq:panic_san_hseqmat_explicit_odd}\\
	\mu_{2n+2} & = & \Delta^n (\mu_2-v)+v, \label{eq:panic_san_hseqmat_explicit_even}
\end{eqnarray}
where $v=h_\mathrm{init}[\begin{array}{cc}1 & 1\end{array}]^T$.

We can simplify Eq.(\ref{eq:panic_san_hseqmat_explicit_odd}) and Eq. (\ref{eq:panic_san_hseqmat_explicit_even}) further to obtain:
\begin{eqnarray}\label{eq:panic_san_hqseq_explicit}
	h_{2n+1} & = & \frac{d}{2q}(h_1-2h_0+h_\mathrm{init})\left(\lambda_1^n+\lambda_2^n\right) + \frac{1}{2}(h_1-h_\mathrm{init})\left(\lambda_1^n-\lambda_2^n\right)+ h_\mathrm{init} \\
	h_{2n+2} & = & \frac{d}{2q}(h_1-2h_1+h_\mathrm{init})\left(\lambda_1^n+\lambda_2^n\right) + \frac{1}{2}(h_2-h_\mathrm{init})\left(\lambda_1^n-\lambda_2^n\right) + h_\mathrm{init},
\end{eqnarray}

where $\lambda_1=\frac{1}{2}d(d-2+q)$ and $\lambda_2=\frac{1}{2}d(d-2-q)$, with $q=\sqrt{d^2-4d}$ are the eigenvalues of $\Delta$ in Eq. (\ref{eq:panic_san_hseqmat}).

Therefore, under the assumptions we made in the beginning of this section, there are the following possibilities regarding the fate of node $i$:

\begin{enumerate}
	\item If $d>4$, $\lambda_1, \lambda_2 \in \mathbb{R}$ and $|\lambda_1|=\lambda_1>4>1$. Thus, the system is unstable and $h_{i,t}$ will tend to infinitely large negative values after node $i$ becomes disconnected from the network. Consequently, the probability of a new link involving node $i$ tends to $0$ exponentially quickly. Moreover, $\lim_{d\rightarrow\infty}\lambda_1=\infty$ and $\lim_{d\rightarrow\infty}\lambda_2=1^+$.
	\item If $0<d<1$, $\lambda_1=\lambda_2=d$. Therefore $|\lambda_1|<1$ and $|\lambda_2|<1$. Thus, the system is stable and $h_{i,t}$ will eventually return to values close to $h_{i,0}$, which allow for the creation of links between node $i$ and the rest of the network.
	\item If $1<d<4$, the evolution of $h_{i,t}$ would be unstable and unbounded for $\tau>1$ in the absence of the asymmetry in the panic factor defined in Eq. (\ref{eq:model_specs_h}). However, this asymmetry condition gives rise to a situation in which $h_{i,t}$ eventually returns to a point close to $h_{i,0}$, where link formation is possible. This happens when $\delta h_{i,t}$ becomes non-negative, which implies $d\cdot\min\left(0,\ \delta h_{i,t}\right)=0$.
\end{enumerate}

\begin{figure}[!htb]
	\centering
	\includegraphics[scale=0.45]{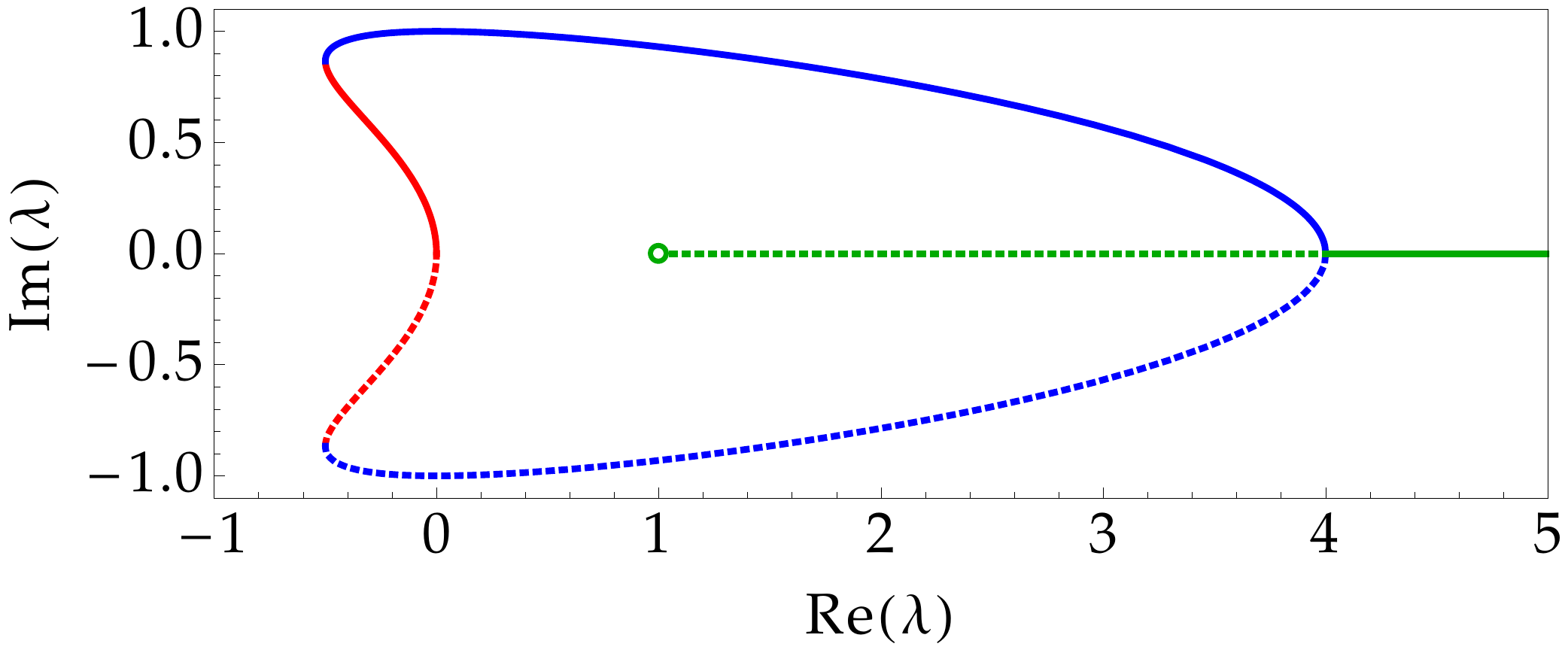}
	\caption[Content]{Parametric representation of $\lambda_1$ and $\lambda_2$ in the complex plane. The unstable case ($d>4$) is depicted in green, while the stable regime ($0<d<1$) is in red. The case with $1<d<4$, in which there is instability but $h_{i,t}$ is bounded, is in blue.}
	\label{fig:panic_ev}
\end{figure}

The eigenvalues $\lambda_1$ and $\lambda_2$ corresponding to the cases above are represented in Fig. \ref{fig:panic_ev}. The interested reader is referred to \cite{Batista2015} for further details.

\bibliography{references}

\end{document}